\documentclass[twocolumn]{aastex631}
\usepackage{graphicx,subfigure,amsmath,amsfonts,amssymb,multirow,extarrows,bm}

\begin{document}
\newcommand{\PMO}{Key Laboratory of Dark Matter and Space Astronomy, Purple Mountain Observatory, Chinese Academy of Sciences, Nanjing, 210033, People's Republic of China.}
\newcommand{\USTC}{School of Astronomy and Space Science, University of Science and Technology of China, Hefei, Anhui 230026, People's Republic of China.}

\title{Nonparametric Representation of Neutron Star Equation of State Using Variational Autoencoder}
\author[0000-0001-9034-0866]{Ming-Zhe Han}
\author[0000-0001-9120-7733]{Shao-Peng Tang}
\author[0000-0002-8966-6911]{Yi-Zhong Fan}
\email{Corresponding author: yzfan@pmo.ac.cn}
\affiliation{\PMO}
\affiliation{\USTC}
\date{\today}

\begin{abstract}
We introduce a new nonparametric representation of the neutron star (NS) equation of state (EoS) by using the variational autoencoder (VAE).
As a deep neural network, the VAE is frequently used for dimensionality reduction since it can compress input data to a low-dimensional latent space using the encoder component and then reconstruct the data using the decoder component.
Once a VAE is trained, one can take the decoder of the VAE as a generator.
We employ 100,000 EoSs that are generated using the nonparametric representation method based on \citet{2021ApJ...919...11H} as the training set and try different settings of the neural network, then we get an EoS generator (trained VAE's decoder) with four parameters.
We use the mass\textendash{}tidal-deformability data of binary neutron star (BNS) merger event GW170817, the mass\textendash{}radius data of PSR J0030+0451, PSR J0740+6620, PSR J0437-4715, and 4U 1702-429, and the nuclear constraints to perform the joint Bayesian inference.
The overall results of the analysis that includes all the observations are $R_{1.4}=12.59^{+0.36}_{-0.42}\,\rm km$, $\Lambda_{1.4}=489^{+114}_{-110}$, and $M_{\rm max}=2.20^{+0.37}_{-0.19}\,\rm M_\odot$ ($90\%$ credible levels), where $R_{1.4}$/$\Lambda_{1.4}$ are the radius/tidal-deformability of a canonical $1.4\,\rm M_\odot$ NS, and $M_{\rm max}$ is the maximum mass of a non-rotating NS.
The results indicate that the implementation of the VAE techniques can obtain the reasonable results, while accelerate calculation by a factor of $\sim$ 3\textendash10 or more, compared with the original method.
\end{abstract}

\section{Introduction}
A neutron star (NS) has physical conditions that we can hardly achieve in terrestrial experiments. It can allow us to study the behavior of dense matter under extreme conditions (see Refs \citealt{2012ARNPS..62..485L, 2016PhR...621..127L, 2016ARA&A..54..401O, 2017RvMP...89a5007O,Lattimer:2021emm} for reviews).
The states of matter in a stable NS can be described using the so-called equation of state (EOS), i.e., the relationship between the pressure and the energy density at zero temperature.
In low- and very high-density regions, the EOS is well understood \citep{2018RPPh...81e6902B}, while between the two regions, there remain uncertainties.

Up to now, the NICER collaboration has reported two mass\textendash{}radius ($M-R$) measurements of the isolated NS PSR J0030+0451 \citep{2019ApJ...887L..21R, 2019ApJ...887L..24M} and the massive NS PSR J0740+6620 \citep{2021ApJ...918L..27R,2021ApJ...918L..28M}.
These two measurements have been used to constrain the NS EOS in many works \citep{2019ApJ...887L..25B,2019ApJ...887L..26B,2019ApJ...887L..22R,2020ApJ...892...55J,2021ApJ...918L..29R,2021ApJ...914L..15B,2021PhRvD.104f3032T}.
In addition to the $M-R$ measurements, the well-known gravitational wave (GW) event from the binary neutron star (BNS) merger GW170817 \citep{2017PhRvL.119p1101A,2019PhRvX...9a1001A}, which can be used to calculate the tidal deformability ($\Lambda$) of the NSs, has also inspired many studies about the NS EOS \citep{2018PhRvL.121p1101A,2018PhRvL.120q2703A,2018PhRvL.120q2702F,2018ApJ...868L..22L,2018PhRvL.121f2701L,2018PhRvL.120z1103M,2019PhRvD..99l3026K,2019ApJ...885...39J}.
The phenomenological methods are commonly used for extracting information from various observations, which can be further divided into two categories, i.e., the parametric and nonparametric methods.
The parametric method, for instance, the spectral expansion \citep{2010PhRvD..82j3011L} and the piecewise polytropes \citep{2009PhRvD..79l4032R, 2016ApJ...820...28O, 2017ApJ...844..156R}, have been proved to be useful in constraining NS EOS.
However, the parametric method may significantly rely on the parametric form, resulting in a biased outcome due to misspecification.
Therefore, we need a method that does not depend on a specific parametric form, i.e., the nonparametric method.
The Gaussian process (GP) has been used as a nonparametric method \citep{2019PhRvD..99h4049L,2020PhRvD.101f3007E,2020PhRvD.101l3007L}, but such a method is not easy to be incorporated by Bayesian inference with the Markov Chain Monte Carlo (MCMC) algorithm due to the nontrivial jump proposals\citep{titsias_rattray_lawrence_2011}.
In \citet{2021ApJ...919...11H}, we developed a nonparametric method via the feed-forward neural network (FFNN), and by using the sampling algorithm {\sc MultiNest} we obtained the posterior distributions of EOS using the NS observations.
To make the model nonparametric, we had to use 31 parameters in the FFNN; thus, the nonparametric method has far more parameters than the parametric method, which may increase the calculation cost and make the sampling algorithm hard to converge.

Deep learning has recently become a powerful method in astrophysical data analysis.
\citet{2018PhRvD..98b3019F,2020PhRvD.101e4016F,Fujimoto:2021zas} have developed a supervised learning method to constrain the NS EOS, where they used piecewise polytropes to represent the EOS. They took the squared sound speed $c_{\rm s}^2$ at corresponding pressure as the output of the network, and the mass, radius, and their variances as the input.
Therefore, one can use the NS observations to get the parameters of the NS EOS via the trained network.
Besides, \citet{2022arXiv220101756S} have trained two networks, one is for generating the EOS (EOS Network), and the other one is trained to solve Tolman–Oppenheimer–Volkoff (TOV) equations (TOV-Solver Network), i.e., translate the EOS ($p(\rho)$) to the NS observations ($M(R)$ and $M(\Lambda)$ curves). Then the authors take the difference between the predicted quantities and the real observations as the loss function to train the network.
Once the loss function converges, they can use the EOS Network to generate the desired EOS.
However, both the above two methods are deterministic, and they estimate the uncertainties by just repeating the optimization procedure many times.
As mentioned in the previous paragraph, the nonparametric method introduced in \citet{2021ApJ...919...11H} combines the nonparametric representation of the EOS and the Bayesian inference, which can naturally handle the uncertainties. However, the high dimensionality of the parameters in such a method increases the difficulty of sampling.

In deep learning, the variational autoencoder (VAE) is a generative neural network that is commonly used for dimensionality reduction.
The VAE and the other variants based on it have also been widely used in astronomy \citep{2020PhRvD.102j4057G,2022PhRvD.106h3022B,2022PhRvD.105l4021W,2022NatPh..18..112G,2022AJ....164..263M}.
In this work, we use the VAE to reduce the dimension of the parameter space in the nonparametric representation and use the Bayesian method to obtain the posterior distributions of the NS EOS parameters given the NS observations.
In Sec.~\ref{sec:method} we first review the nonparametric representation of the NS EOS in \citet{2021ApJ...919...11H} and then introduce the architecture of the VAE and the training process.
We summarize the observation data used in this work in Sec.~\ref{sec:observs} and present the results in Sec.~\ref{results}.
Finally, we give the summary and discussion in Sec.~\ref{summary}.

\section{Method}\label{sec:method}
\subsection{Feed-forward neural network}\label{ffnn}
In our previous work \citep{2021ApJ...919...11H}, we introduced a nonparametric representation of NS EOS. Here we briefly recall the method in \citet{2021ApJ...919...11H}.
In that work, the NS EOS can be described by an FFNN model with a single hidden layer,
\begin{equation}
    \phi = \sum^N_{i=1} w_{2i} \sigma (w_{1i} \log p + b_{i}) + B, 
\label{eq:FFNN}
\end{equation}
where $\phi$ is an auxiliary variable defined as 
\begin{equation}
    \phi = \log \left(c^2 \frac{\mathrm{d}\varepsilon}{\mathrm{d}p} - 1 \right).
\end{equation}
In the above equations, $p$ is the pressure, $\varepsilon$ is the energy density, $w_{1i}$/$w_{2i}$ are the weights parameters, $b_{i}$ ($B$) are bias parameters (B can also be considered as the overall residual), $N$ is the number of the neural nodes (or the width of the network), and $\sigma(\cdot)$ stands for a nonlinear function (the so-called activation function).
The activation function we choose here is the sigmoid function, 
\begin{equation}
    \sigma(x) = \frac{1}{1 + e^{-x}},
\end{equation}
which can guarantee the requirement of sigmoidal functions (for more details, see \citealt{Cybenko1989} and \citealt{2021ApJ...919...11H}), i.e., $\sigma(x) \rightarrow 0(1)$ when $x \rightarrow -\infty(+\infty)$.
In this work, we use a slightly different version of that model, which reads 
\begin{equation}
    c_s^2/c^2 = \sigma\left(\sum^N_{i=1} w_{2i} \sigma (w_{1i} \log \rho + b_{i}) + B \right).
\label{eq:ffnn_new}
\end{equation}
We take the rest-mass density $\rho$ as the input variable and the squared sound speed $c_{\rm s}^2/c^2={\rm d}p/{\rm d}\varepsilon$ as the output variable, instead of $\phi$ and $p$.
One can easily find that $\sigma(-\phi)=c_s^2/c^2$, so using the squared sound speed with a sigmoid activation function as the output is almost the same as using the auxiliary variable $\phi$ as the output.
As for the input, the rest-mass density is more straightforward for applying the multiple constraints, e.g., the nuclear constraints that directly constrain the pressures at specific rest-mass densities.
These variations in the input/output variables only have a little effect on the results, which could be negligible.
Besides, there is another choice of the activation function in Eq. (\ref{eq:ffnn_new}), the hyperbolic tangent function. 
The differences between these two activation functions have been discussed in the Appendix A of \citet{Han:2022rug}, and we do not discuss the influence of activation functions in this work.
With the FFNN in hand, we can now make a representation of the NS EOS.

\subsection{Variational AutoEncoder}\label{vae}
\begin{figure*}[ht]
    \centering
    \includegraphics[width=0.9\textwidth]{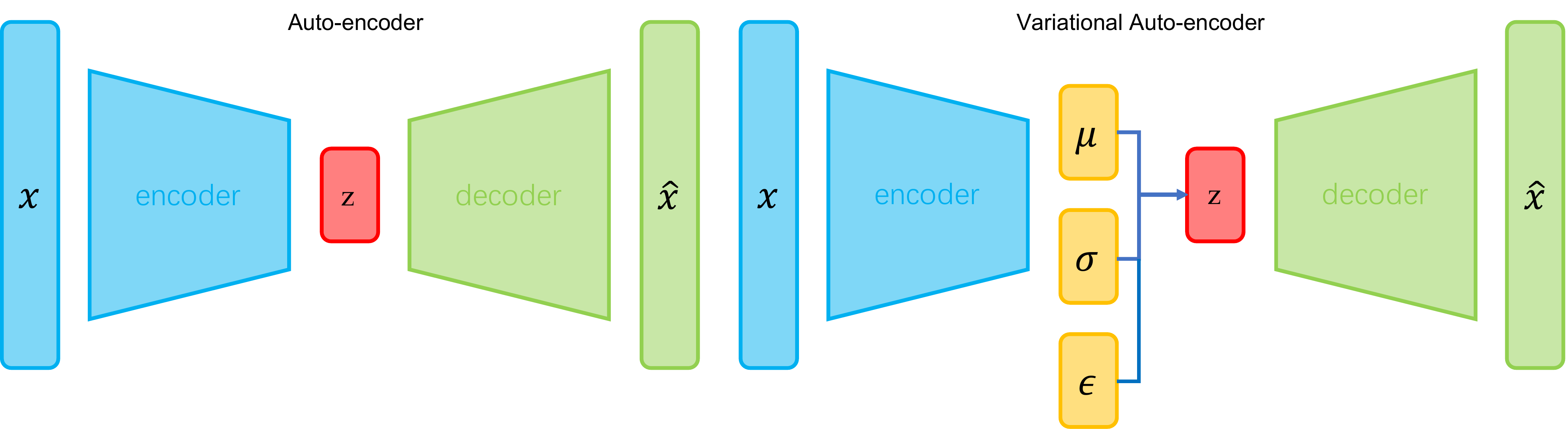}
    \caption{Structure of the two neural networks. The left panel is the autoencoder, while the right is the variational autoencoder. $x$ denotes the data, $\hat{x}$ is the reconstruction of the data $x$, and $z$ is the latent variable in latent space $Z$. In the right panel, $\mu$ and $\sigma$ are the mean and standard deviation of the latent variable $z$, and $\epsilon$ is a random variable that follows the standard normal distribution.}
    \label{fig:networks}
    \hfill
\end{figure*}
The VAE is a deep generative model, and it is very similar to the autoencoder (AE), so we briefly introduce the AE first.
A typical AE structure is shown in the left panel of Fig.~\ref{fig:networks}, which consists of two parts: the encoder and the decoder.
The goal of an encoder is to learn a mapping from the data $x$ to a low-dimensional latent space $Z$, where the reduction of dimensionality can help us to compress the data and get a compact feature representation of them.
And note that this is an unsupervised learning problem since we do not have labels in the training set.
Therefore, by minimizing the reconstruction error between $x$ and $\hat{x}$, e.g., the mean square error $L(x, \hat{x})=\|x-\hat{x}\|^2$, we can learn the latent representation of the data by itself without any labels (that is why we call it AE, i.e., automatically encoding data).
Nevertheless, there is a shortcoming in the AE. The encoder of AE is deterministic, which means that if we draw a random sample of the latent vector and put it into the decoder, we may not be able to get the desired result (i.e., a new sample that is not in the training set but similar to those in the training set).

The fundamental distinction between the VAE and AE models is that the VAE is now a probabilistic model rather than a deterministic one.
The structure of VAE is shown in the right panel of Fig.~\ref{fig:networks}, where we can see that the deterministic layer of AE's encoder is replaced by the so-called sampling layer, i.e., we compute the mean $\mu$ and the standard deviation $\sigma$ from the encoder and draw a random sample $\epsilon$ from the standard normal distribution, then the latent variable $z$ is computed by: $\vec{\mu} + \vec{\sigma} \odot \epsilon$.
Therefore, the goal of the VAE can be described in a probabilistic manner as follows: the encoder is going to be trained to compute the probability distribution of the latent variable $z$ given the input data $x$, i.e., $q_{\phi}(z|x)$, while the decoder is going to take that learned latent representation and compute a new probability distribution of the input data $x$ given the latent distribution of $z$, i.e., $p_{\theta}(x|z)$.
However, computing the $q_{\phi}(z|x)$ analytically is impossible due to its high dimensionality, and using a numerical method like MCMC is too expensive in computation, so usually, we use a prior distribution to approximate the target distribution, i.e., the variational inference (VI).
Thus, for approximating the target distribution, we need to reduce the difference between $q_{\phi}(z|x)$ and the prior $p(z)$. The difference between two distributions ($Q(x)$ and $P(x)$) is usually measured by the KL divergence $D_{\rm KL}$, which is defined by
\begin{equation}
    D_{\rm KL}(Q(x)\|P(x)) = \int Q(x)\log \frac{Q(x)}{P(x)}dx.
\label{eq:Dkl}
\end{equation}
Now we take a look at the loss function of the VAE,
\begin{equation}
    L_{\rm total} = L_{\rm rec} + L_{\rm KL} = \|x-\hat{x}\|^2  - \frac{1}{2}\sum_{\rm j=0}^{\rm k-1}(\mu_{\rm j}^2 + \sigma_{\rm j} - 1 - {\rm log}(\sigma_{\rm j})).
\label{eq:loss_vae}
\end{equation}
The total loss function has two terms, the reconstruction error $L_{\rm rec}$ and the regularization term $L_{\rm KL}$. The first term is just like the reconstruction error in AE, and the second term is the KL-divergence between the prior distribution $p(z)$ (here the prior distribution is the standard normal distribution) and the target distribution $q_{\phi}(z|x)$.
The reason why we call $L_{\rm KL}$ the regularization term is that it can encourage the encodings to distribute evenly in the center region of the latent space, and punish the network when it tries to ``cheat" by clustering the points in specific regions (i.e., without the regularization term, the output deviation of the encoder $\sigma$ is almost zero and the VAE degenerates to AE).

\subsection{Training process}
\label{sec:train}
\begin{table}[ht]\footnotesize
    \begin{center}
    \renewcommand\arraystretch{1}
    \tabcolsep=0.1cm
        \begin{tabular}{lccc}
            \hline \hline
            Layer & Type & Number of Neurons & Activation Function \\
            \hline
            Input & - & 128 & - \\
            Layer 1 & Dense & 64 & ReLu \\
            Layer 2 & Dense & 64 & ReLu \\
            Layer 3 & Dense & 32 & ReLu \\
            Layer 4 & Dense & 32 & ReLu \\
            Latent Layer & Lambda & 4 & - \\
            Layer 5 & Dense & 32 & ReLu \\
            Layer 6 & Dense & 32 & ReLu \\
            Layer 7 & Dense & 64 & ReLu \\
            Layer 8 & Dense & 64 & ReLu \\
            Output & Dense & 128 & Sigmoid \\
            \hline \hline
        \end{tabular}
    \end{center}
    \label{tab:ffnn}
    \caption{Hyper parameters of the VAE we used in this work. The latent layer consists of two parts: the first part is made of two dense layers that take the output of layer $4$ as input, and output the mean and standard deviation of a multivariate normal distribution, and the second part is the so-called sampling layer that draws samples from a multivariate normal distribution whose mean and standard deviation are computed by the first part.}
\end{table}

This section aims to train a VAE decoder, i.e., the EOS generator, which can generate EOSs from a low-dimensional latent space.
Before the training, we need to generate the training set of the NS EOS.
Note that the training set is just the set of the NS EOSs, which can be any physically realistic NS EOSs.
In this work, we use the FFNN model based on \citet{2021ApJ...919...11H} to generate the training set.
We randomly draw the samples of the FFNN parameters, i.e., $w_{1i}$, $w_{2i}$, $b_i$, and $B$ (31 in total, and each parameter is uniformly sampled in (-5, 5)), and use the FFNN model to calculate the corresponding EOS.
Once we get the EOS, we need to solve the TOV equations with the EOS, and this process is too complicated to be done analytically.
Therefore, we usually solve it numerically with a tabulated EOS. The resolution of the EoS table is 128D, namely an NS EOS is represented by a 128D $c_s^2(\rho)$ array, which is controlled by the 31 FFNN parameters.
The rest-mass densities of tabulated points are logarithmically uniform in [$\sim 0.3\rho_{\rm sat}, 10\rho_{\rm sat}$].
We sample 100,000 EoSs from the priors, whose maximum masses satisfy the condition of $M_{\rm max} \in (1.4, 3)~M_{\odot}$.

\begin{figure}[ht]
    \centering
    \includegraphics[width=0.95\columnwidth]{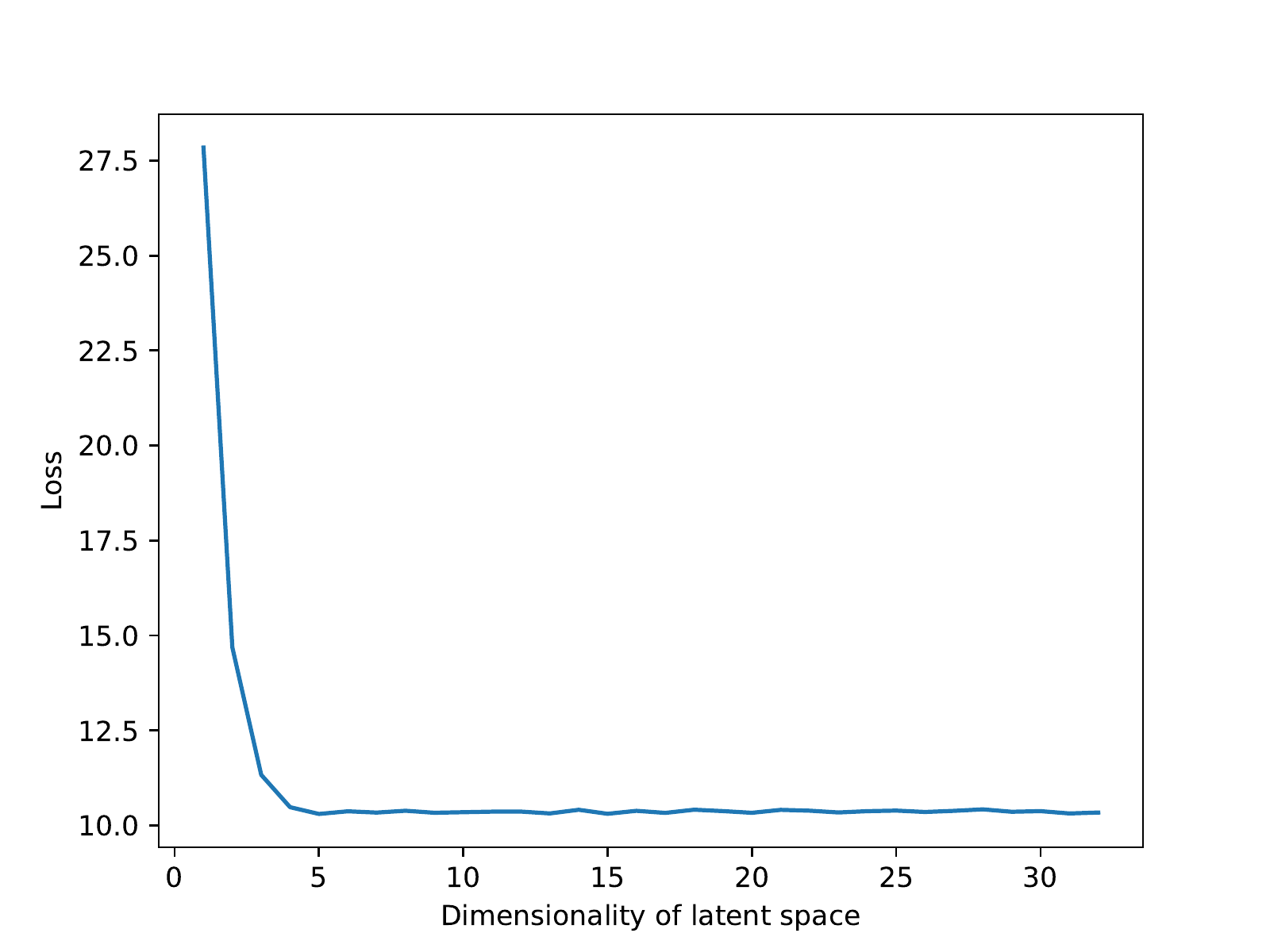}
    \caption{Losses after training for 200 epochs with different settings for the dimensionality of latent space.}
    \label{fig:loss}
    \hfill
\end{figure}

After the training set has been generated, we can then use it to train the neural network.
We build a VAE neural network, whose architecture is shown in Tab.~\ref{tab:ffnn}.
We use Python package Keras \citep{2018ascl.soft06022C} in TensorFlow \citep{2016arXiv160508695A} and the Adam \citep{2014arXiv1412.6980K} optimizer, with a learning rate of $0.0001$ and a batch size of $32$.
By minimizing the loss function defined in Eq. (\ref{eq:loss_vae}), we can get the trained VAE model.
As for the choice of dimensionality of the latent space, we test several settings (1-32), and for each setting, we train the model for 200 epochs.
From Fig.~\ref{fig:loss}, we can see that at low dimensions (1-4), the loss of the model decreases rapidly as the dimension of the latent space increases; when the dimension of the latent space is larger than 4, the loss of model converges.
Therefore, it is reasonable to use a 4D latent space in this work.
Finally, once the VAE neural network has been trained, we can draw a random vector from the 4D standard normal distribution and then use the trained decoder to reconstruct the 128D EOS table.

To summarize, we first use the FFNN model to generate the training set, which is to convert the 31D joint uniform distribution of the FFNN parameters, i.e., $w_{1i}$, $w_{2i}$, $b_i$, and $B$, to the 128D joint distribution of the EOS parameters, i.e., the squared sound speed at corresponding rest-mass densities (128D) mentioned in the previous paragraph.
We then train a VAE model with the training set, and take out the trained decoder part.
This process is to convert the above 128D joint distribution of the EOS parameters ($c_s^2(\rho)$) to the 4D joint standard normal distribution of the latent variables, i.e., $z_1$, $z_2$, $z_3$, and $z_4$.
As a result, we transform the nonparametric representation's prior of the NS EOS from a 31D uniform joint distribution to a 4D joint standard normal distribution. At the same time, it still contains the degrees of freedom of the model (compared to the parametric models).
Now we can use just $4$ parameters to control an NS EOS.

\section{Observations}\label{sec:observs}
Complementary to terrestrial nuclear experiments, NS observations can be used to constrain the EOS of matter at supranuclear density.
Recently, the radius measurements of PSR J0740+6620, the heaviest pulsar known, have been obtained by the scientific team of NICER.
With the extra information from radio timing \citep{2021ApJ...915L..12F} and {\it XMM-Newton} spectroscopy, the radius of this massive NS was inferred by the pulse profile modeling of the hotspot's light-curve, which is ${12.39}_{-0.98}^{+1.30}\,{\rm km}$ (by \citealt{2021ApJ...918L..27R}) or ${13.7}_{-1.5}^{+2.6}\,{\rm km}$ (by \citealt{2021ApJ...918L..28M}) at $68\%$ credible level.
In comparison to the first results obtained by NICER in 2019, i.e., the simultaneous mass-radius measurement of the isolated NS PSR J0030+0451 \citep{2019ApJ...887L..21R, 2019ApJ...887L..24M}, the massive pulsar PSR J0740+6620 shares almost the same radius with PSR J0030+0451, though their masses differ $>50\%$ from each other.
Since more massive NS generally has a larger central density, such measurements allow us to probe the EOS at densities much higher than those based on previous NS observations.
Meanwhile, the very nearby pulsar PSR J0437-4715, whose mass ($\sim 1.44\,M_\odot$) is determined by the reliable timing analyses \citep{2016MNRAS.455.1751R}, is one of the prime targets for NICER \citep{2019ApJ...887L..27G}.
The radius of this object has been updated in \citet{2019MNRAS.490.5848G} and will be directly tested by the dedicated NICER observations in the near future.
Besides, via the so-called cooling-tail method, the mass$-$radius measurement of 4U 1702-429 was obtained with small uncertainty by \citet{2017A&A...608A..31N}. The tidal deformability measurement of the landmark event GW170817 \citep{2017PhRvL.119p1101A, 2019PhRvX...9a1001A}, originating from the merger of two NSs, has also given us a large opportunity to study the EOS \citep{2018PhRvL.121p1101A}.
Therefore, we use all of the observation data discussed above to perform joint analysis, which includes the tidal-deformability measurements from GW170817, and the mass$-$radius measurements of PSR J0030+0451, PSR J0740+6620, PSR J0437-4715, and 4U 1702-429.
To simplify, we use $\mathcal{D}_1$ to stand for the data set containing GW170817, PSR J0030+0451, and PSR J0740+6620 data, and $\mathcal{D}_2$ to denote the data set that contains all of the observation data discussed above.

Supposing that all NSs share the same EOS, we can take the following likelihood
\begin{eqnarray}\label{eq:likelihood}
    \mathcal{L} = & \mathcal{L}_{\rm GW}(d\mid\vec{\theta}_{\rm GW})&\times \prod_{i} \mathcal{P}_i(M(\vec{\theta}_{\rm EOS}, h_i), R(\vec{\theta}_{\rm EOS}, h_i)) \nonumber \\
    &\times \mathcal{L}_{\rm Nuc}(\vec{\theta}_{\rm EOS})&
\end{eqnarray}
to constrain the EOS parameters $\vec{\theta}_{\rm EOS}$ by performing Bayesian inference with {\sc Bilby} \citep{2019ascl.soft01011A} and {\sc PyMultiNest} \citep{2014A&A...564A.125B} packages.
The $\mathcal{L}_{\rm Nuc}(\vec{\theta}_{\rm EOS})$ is the likelihood of the nuclear constraints, which have also been used in \citet{2021ApJ...919...11H}.
The likelihood is 1 when all the nuclear constraints are satisfied, else 0, where the nuclear constraints are $3.12\times10^{33} ~{\rm dyn ~cm^{-2}}\leq p(\rho_{\rm sat})\leq 4.70\times10^{33}~{\rm dyn ~ cm^{-2}}$ \citep{2014EPJA...50...40L,2017ApJ...848..105T,2019ApJ...885...39J} and $p(1.85\rho_{\rm sat})\geq 1.21\times10^{34}~{\rm dyn ~cm^{-2}}$ \citep{2016ApJ...820...28O}.
Since the results obtained by using data of \citet{2019ApJ...887L..21R}/\citet{2021ApJ...918L..27R} and \citet{2019ApJ...887L..24M}/\citet{2021ApJ...918L..28M} are nearly consistent with each other \citep{2021PhRvD.104f3032T}, we only use the data of \citet{2019ApJ...887L..21R} for PSR J0030+0451\footnote{The data of ST+PST case is considered, see \url{http://doi.org/10.5281/zenodo.3386449}} and \citet{2021ApJ...918L..27R} for PSR J0740+6620\footnote{The data file ``STU/NICERxXMM/FI\_H/run10" from \url{https://doi.org/10.5281/zenodo.4697625} is taken into account.}. For GW170817, we take the interpolated marginalized likelihood from \citet{2020MNRAS.499.5972H} into analysis, which shows good consistency with the original GW data. For mass-radius measurements, we use the Gaussian kernel density estimation of the publicly distributed posterior samples of mass and radius to build the likelihood (see \citet{2021PhRvD.104f3032T} for more details).

\section{Results}\label{results}
\begin{figure*}[ht]
\centering
\includegraphics[width=0.9\textwidth]{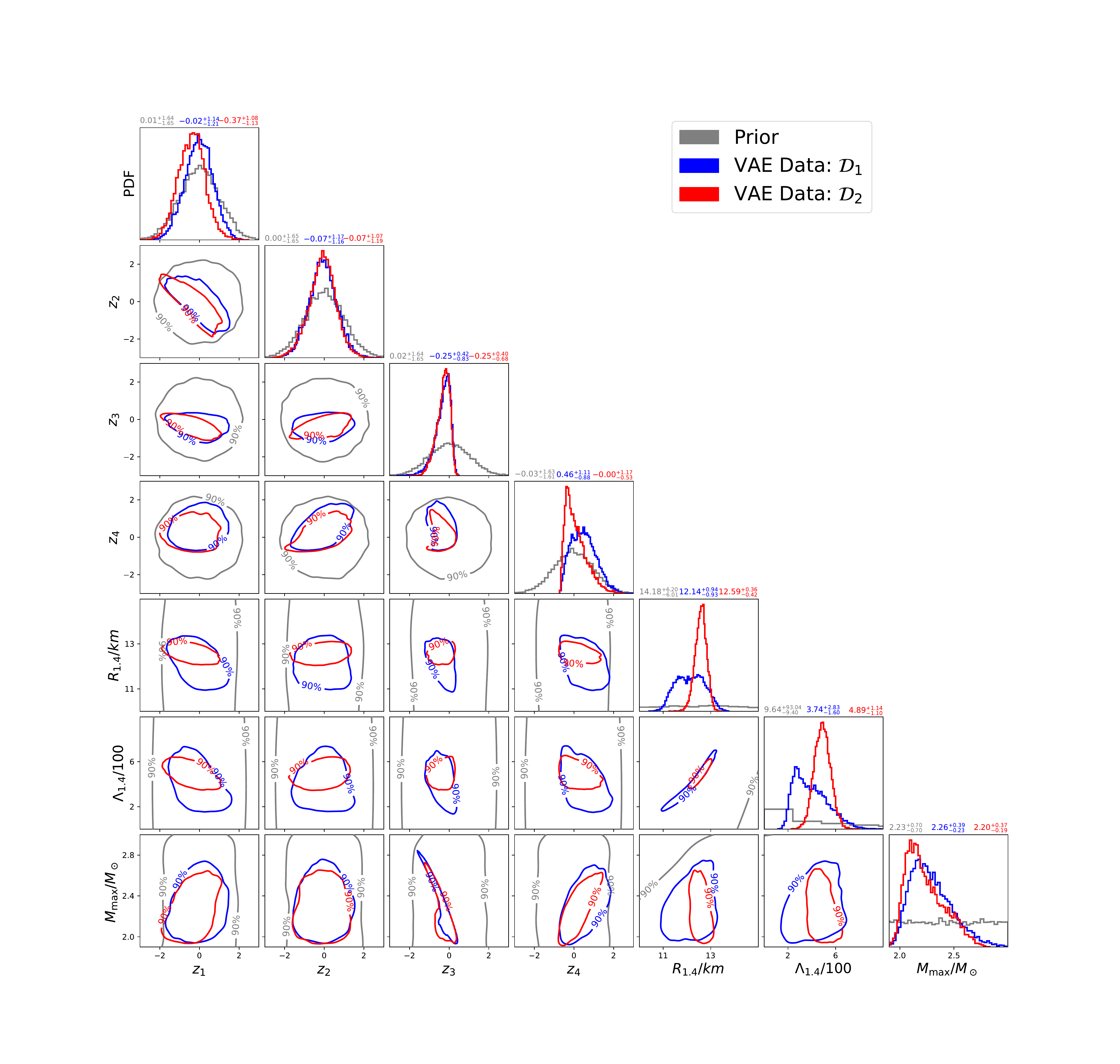}
\caption{
Posteriors of the latent variables and bulk properties of NS.
The red curves and blue curves are the results of the data set $\mathcal{D}_1$ and $\mathcal{D}_2$, respectively.
The gray curves are the priors of the latent variables, which are all standard normal distributions. The priors of bulk properties of NS are almost uniform.
All the results are at the $90\%$ credible level.
}
\label{fig:lvpost}
\hfill
\end{figure*}
After using the VAE to represent the NS EOS, the EOS can be described by only $4$ parameters. The direct results of the Bayesian inference are the posteriors of the latent variables, i.e., $z_1$, $z_2$, $z_3$, and $z_4$.
These latent variables are also called hidden variables because they do not have any direct relations to the reconstructed data. Thus, we need to convert the above results (latent variables) to data (the EOS tables) that we can understand using the trained VAE decoder, i.e., decoding.
The posterior distributions of the latent variables and the bulk properties of NS are shown in Fig.~\ref{fig:lvpost}.
We find that the radius and tidal deformability of a canonical $1.4 M_\odot$ NS have a strong correlation, and the data set $\mathcal{D}_1$ ($\mathcal{D}_2$) gives $R_{1.4}=12.14^{+0.94}_{-0.93}\,\rm km$ ($12.59^{+0.36}_{-0.42}\,\rm km$) and $\Lambda_{1.4}=374^{+283}_{-160}$ ($489^{+114}_{-110}$) at $90\%$ credible level.
Except for the parameter $z_2$, the other three latent variables' posteriors are obviously different from their priors, which means that the observation data are informative.
And for the variable $z_4$, the results of data sets $\mathcal{D}_1$ and $\mathcal{D}_2$ show a little difference, while for the other three latent variables, the results have no apparent discrepancies between the two data sets.
Interestingly, the variables $z_1-z_2$ show a correlation in the joint distribution. And the variables $z_3$ and $z_4$ are correlated with the maximum mass of nonrotating NS.
The maximum mass $M_{\rm max}$ is constrained to be $M_{\rm max}=2.26^{+0.39}_{-0.23} \rm M_\odot$ ($2.20^{+0.37}_{-0.19} \rm M_\odot$) for data set $\mathcal{D}_1$ ($\mathcal{D}_2$), which is consistent with some previous results \citep{2020PhRvD.102f3006S,2021PhRvD.104f3032T, 2021ApJ...908L..28N}.

\begin{figure*}[ht]
\centering
\includegraphics[width=0.9\textwidth]{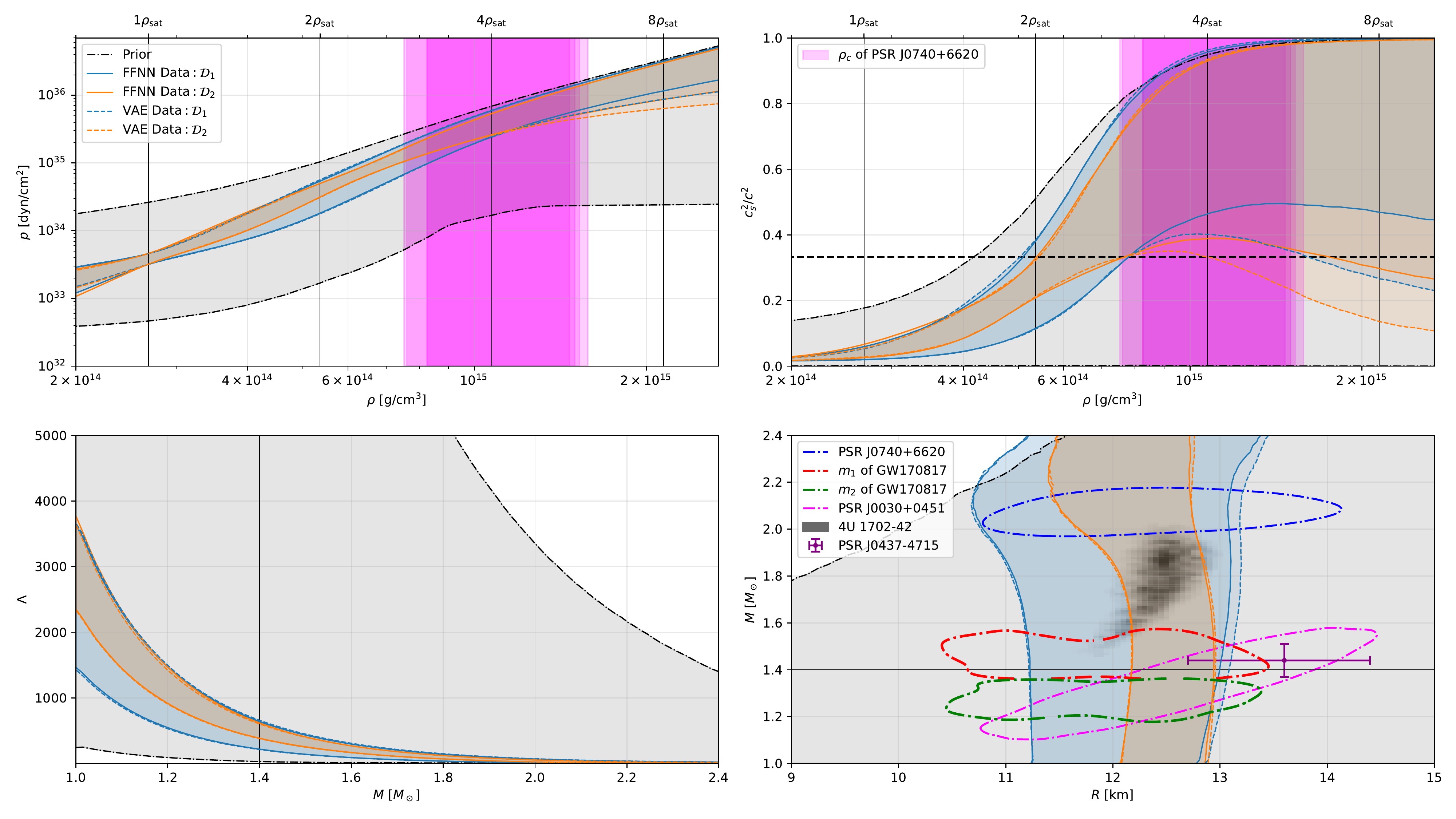}
\caption{
Posterior distributions for the pressure (p) vs. the rest-mass density ($\rho$) (upper left), the squared speed of sound divided by the squared light speed in vacuum $c_{\rm s}^2/c^2$ vs. $\rho$ (upper right), the dimensionless tidal deformability ($\Lambda$) vs. mass (M) (lower left), and M vs. the radius (R) (lower right).
All of the uncertainty regions are at the $90\%$ credible level.
The prior is shown by black dashed-dotted lines. The result of using data set $\mathcal{D}_1$ and $\mathcal{D}_2$ are shown with the blue and orange lines, respectively.
For comparison, we also draw the results obtained by using a similar method to that of \citet{2021ApJ...919...11H} (dashed lines). i.e., the FFNN model.
The black vertical lines in the upper panels denote several typical densities (1$\rho_{\rm sat}$, 2$\rho_{\rm sat}$, 4$\rho_{\rm sat}$, and 8$\rho_{\rm sat}$, where $\rho_{\rm sat}$ is the nuclear saturation density.), and the vertical magenta regions are the central density of PSR J0740+6620.
Besides, the horizontal dashed black line in the panel of $c_{\rm s}^2/c^2(\rho)$ (upper right) is the conformal limit, i.e., $c_{\rm s}^2/c^2=1/3$, while the black straight lines in the lower panels stand for 1.4 $\rm M_\odot$. The $M-R$ measurements (at $68.3\%$ credible level) of PSR J0030+0451, PSR J0740+6620, PSR J0437-4715, and 4U 1702-429 are represented by the blue dotted-dashed contour, magenta dotted-dashed contour, purple error bar, and gray area, respectively. The red and green dotted-dashed contours represent the $M-R$ posteriors of the NS associated with GW170817 (adopted from the right panel of Fig.~3 of \citealt{2018PhRvL.121p1101A}).
}
\label{fig:posterior}
\hfill
\end{figure*}

After the decoding, we can now discuss the results of the EOS directly.
To illustrate the efficacy of the VAE method, we also perform the Bayesian inference directly to the FFNN model, which is controlled by 31 parameters.
The likelihood function is the same for these two methods; thus we can directly compare these results.
In the upper panels of Fig.~\ref{fig:posterior}, we can see that at most densities regimes the results of the FFNN model (solid lines) are almost the same as that of the VAE model (dashed lines). The consistencies of the reconstructed $\Lambda-M$ and $M-R$ (see the lower panels) are even more remarkable. 
This indicates that the VAE approaches have been implemented very successfully because we can utilize the VAE model, which only has four parameters, to produce the same outcomes as the FFNN model, which has 31 parameters, while spending less time.
The use of VAE techniques in this work has the potential to accelerate calculation times by a factor of $\sim$ 3 or more.
With the data set $\mathcal{D}_1$ ($\mathcal{D}_2$), the FFNN model requires $3.87$ ($45.28$) hr, whereas the VAE model requires only $1.27$ ($18.22$) hr. 
All these calculations are performed in one compute node with 128 cores.
In some cases, the enhancement is even more efficient.
Without incorporating the nuclear constraints, for the data set $\mathcal{D}_1$, the calculation of the VAE model can be more than $10$ times faster than the FFNN model.
Nevertheless, we find that in a high-density region (i.e., $\gtrsim 4 \rho_{\rm sat}$), the constraint on the EOS with the VAE model is less ``stringent" than that of the FFNN model (see the upper left panel of Fig.~\ref{fig:posterior}). The same happens in the upper right panel of Fig.~\ref{fig:posterior}.
This difference is likely caused by the lack of effective probe at such high densities. The magenta regions in the upper panels of Fig.~\ref{fig:posterior} represent the central density of PSR J0740+6620, the most massive neutron star that has been accurately measured so far. Clearly, even for such a massive compact object, the central density can only reach $\sim 4\rho_{\rm sat}$, above which the EOS cannot be effectively constrained.
In view of the above facts, we conclude that the VAE model can yield reasonable results efficiently.

\section{Summary}\label{summary}
In this work, based on \citet{2021ApJ...919...11H} we develop a new Bayesian nonparametric method for studying the NS EOS.
We use the deep neural network VAE to reduce the number of parameters that represent the EoS.
By comparing different settings of the network, we find that a VAE with 4D latent space is a proper choice for the representation of EOS.
After the training process, we get a trained decoder network. Then we draw a random vector from a 4D standard normal distribution and use the decoder to convert it to the reconstructed 128D EOS table, i.e., we can represent the NS EOS using only four parameters.
We perform Bayesian inference to infer the EOS posteriors using the NS observations, i.e., the $M-R$ measurements of PSR J0030+0451, PSR J0740+6620, PSR J0437-4715, and 4U 1702-429, as well as the $M-\Lambda$ measurements of GW170817.
Sampling from the posteriors of the latent variables with numerical sampling algorithm {\sc PyMultiNest}, we compute the EOS tables using the trained decoder, and by numerically integrating the TOV equations we finally get the macroscopic properties of NS we are interested in.
The radius and tidal deformability of a canonical $1.4\,\rm M_\odot$ NS are constrained to be $R_{1.4}=12.14^{+0.94}_{-0.93}\,\rm km$ ($12.59^{+0.36}_{-0.42}\,\rm km$) and $\Lambda_{1.4}=374^{+283}_{-160}$ ($489^{+114}_{-110}$) at $90\%$ credible level for data set $\mathcal{D}_1$ ($\mathcal{D}_2$), respectively.
Besides, the maximum mass of a nonrotating NS $M_{\rm max}$ is constrained to be $M_{\rm max}=2.26^{+0.39}_{-0.23}$ ($2.20^{+0.37}_{-0.19} \rm M_\odot$) for data set $\mathcal{D}_1$ ($\mathcal{D}_2$).
As for the latent variables, we find that except for $z_2$, all the latent variables are well constrained, and some show correlations with each other or the macroscopic properties.

Though we use only four parameters to represent the NS EOS with the VAE neural network, it still maintains the nonparametric feature.
This dimensionality reduction process makes a significant development in the Bayesian nonparametric inference of NS EoS because it can dramatically reduce the dimension of the parameter space and effectively reduce the difficulty and time during the sampling.
Quantitively, the VAE techniques can accelerate calculation by a factor of $\sim$ 3-10 or more.
Nevertheless, there are still some aspects to be improved in future work.
As mentioned in Sec.~\ref{results}, the latent variables are hidden variables that do not have any direct relations to the EOS parameters or the macroscopic properties. However, the posteriors obtained in this work do show a few correlations.
Thus, we can further investigate the relationship between the latent variables and the parameters we are interested in, i.e., disentangle the latent variables.
Though we have tried different hyperparameters of the network to find the proper setting of the VAE, the compress process may still lead to the loss of information.
Therefore, in future works, one can still enhance the efficiency of the representation while maintaining accuracy.
Besides, in the low- and very high-density regions, we can also incorporate the constraints from chiral effective field theory \citep{2021PhRvL.127s2701E} and perturbative quantum chromodynamics \citep{2022arXiv220411877G}.

\section{Acknowledgments}
We appreciate the referees for helpful suggestions and thank Dr. J. L. Jiang for the useful discussion and input.
This work was supported in part by NSFC under grants No. 12233011, 11921003 and No. 11525313.

\bibliography{bibtex}

\end{document}